\documentclass[prl,preprintnumbers, twocolumn, superscriptaddress, amsmath, l66amssymb]{revtex4-1}

\usepackage{graphicx}% Include figure files
\usepackage{dcolumn}% Align table columns on decimal pointlip
\usepackage{bm}% bold math：
\usepackage{verbatim}
\usepackage{CJK}
\usepackage{amsbsy} 
\usepackage{microtype}
\usepackage{ulem}
\usepackage{tikz}
\usepackage{braket}
\usepackage{amsmath}

\def\shKeyLab{INPAC and School of Physics and Astronomy, Shanghai Jiao Tong University, Shanghai Key Laboratory for Particle Physics and Cosmology, Shanghai 200240, China}
\def\THU{Dept. of Phys., Tsinghua University, Beijing, 100084, China}
\def\BHU{School of Instrumentation Science and Opto-electronics Engineering, Beihang University, Beijing, 100191, China}
\def\CAEP{Key Laboratory of Neutron Physics, Institute of Nuclear Physics and Chemistry, CAEP, Mianyang, Sichuan 621900, China}
\usepackage{makeidx}
\begin{document}
\title{New Experimental Limits on Exotic Spin-Spin-Velocity-Dependent  Interactions By Using SmCo$_5$ Spin Sources }
\author{Wei Ji}\affiliation{\THU}
\author{Yao Chen}\affiliation{\BHU}
\email[Corresponding author:]{yao.chen@buaa.edu.cn} 
%\affiliation{Dept. of,, U.S.A}
\author{Changbo Fu}\affiliation{\shKeyLab}\email[Corresponding author:]{cbfu@sjtu.edu.cn}
\author{Ming Ding}\affiliation{\BHU}
\author{Jiancheng Fang}\affiliation{\BHU}
\author{Zhigang Xiao}\affiliation{\THU}
\author{Kai Wei}\affiliation{\BHU}
\author{Haiyang Yan}\affiliation{\CAEP}
\date{\today}
\begin{abstract}
We report the latest results of searching for possible new macro-scale spin-spin-velocity-dependent forces 
(SSVDFs)
based on specially designed iron-shielded SmCo$_5$ (ISSC) spin sources and a spin exchange  relaxation free (SERF) co-magnetometer. 
The ISSCs have  high net electron spin densities of about $1.7\times 10^{21}$ cm$^{-3}$, which mean high detecting sensitivity; and low magnetic field leakage of about $\sim$mG level due to iron shielding, which means low detecting  noise. With help from the ISSCs, the high sensitivity SERF co-magnetometer, and the similarity analysis method, 
new constraints on SSVDFs with forms of  $V_{6+7}$, $V_8$, $V_{15}$, and $V_{16}$ have been obtained, which represent the tightest limits  in force range of 5 cm -- 1 km  to the best of our knowledge.
\end{abstract}
\maketitle
%\tableofcontents 

%=================================================================
%=================================================================
%=================================================================
%\section{Introduction}
\paragraph{Introduction--}
%=================================================================
% =================================================================
%=================================================================
Many new light bosons, such as  axion \cite{axion-wilczek.1978, axion-weinberg.1978, axion-PhysRevLett.38.1440}, dark photon \cite{darkPhoton.2010, Dark-Photon-AN2015331}, paraphoton \cite{paraphoton-PhysRevLett.94.151802}, familon and majoron \citep{PDG16}, have been introduced by theories beyond the Standard Model.
If they exist,  these kinds of new bosons may mediate new types of  long-range fundamental forces, or the so-called 5th forces. 
These possible new forces may break the C, P, or T (or their combinations) symmetry \cite{axion-PhysRevLett.38.1440}, and they have been suspected to be answers to questions like the strong CP violation problem \cite{axion-PhysRevLett.38.1440}. 
The possibility of the existence of 5th forces  has been extensively investigated  experimentally \cite{Axion-Review-2010ARNPS,HILL1988253}. 
Many forms of technology have been used to  search for these 
long-range spin- and/or velocity-dependent forces, including
the torsion balance \cite{ritter1990experimental, Torsion-Balance2006PRL,terrano2015short,hammond2007new}, 
the resonance spring \cite{Loong-Nature2003, Loong-PRD2015}, 
the spin exchange relaxation free  (SERF) co-magnetometer \cite{Romalis-PRL2009,heckel2013limits,terrano2015short,wineland1991search},
 nuclear magnetic resonance (NMR) based methods \cite{petukhov2010polarized,yan2015searching,chu2013laboratory}, and other high sensitivity technologies \cite{tullney2013constraints,serebrov2010search,ficek2017constraints,CPT2010a}.

In all of these experiments, in order to increase detecting sensitivities, 
one of the key issues was how to improve the test matter's polarized spin density. 
This is due to the fact that a  Yukawa-like force is proportional to $e^{-r/\lambda}$\cite{16Forms2006Dobrescu}, where r is the source to probe distance and $\lambda$  is the force range. 
For small  $\lambda$, like in the range of about $1<\lambda <100$ cm,  because of  the limited volume $O(\lambda ^3)$ ,   increasing the polarized spin density is  very critical to improve the detecting sensitivity.

It has been pointed out that interactions between two spin-1/2 fermions, which are mediated by spin 0 or spin 1 bosons, could be classified to 16 terms,
 and 9 of them are spin-spin dependent \cite{16Forms2006Dobrescu}. 
 Among the spin-spin dependent terms, 3 of them are static, and 6 of them depend on the relative velocity between two polarized objects. Compared with the static terms \cite{Romalis-PRL2009,heckel2013limits,terrano2015short,wineland1991search,hunter2013using}, the experimental constraints on 
the spin-spin-velocity-dependent forces (SSVDFs) are still rare today \cite{hunter2014using,ficek2017constraints}.
For the latter terms, not only a relative velocity between the source and the probe is required, but also both of them must be  spin-polarized. Therefore, they are more difficult to study experimentally.

In Ref. \cite{ji2017searching}, 
an experimental scheme with high electron spin-density sources,  iron-shielded SmCo$_5$ (ISSCs), was proposed to detect the SSVDFs.
By taking advantage of the high electron spin density of ISSC and the high sensitivity of SERF co-magnetometer \cite{Kornack:2005},
the proposed system had a potential to detect several SSVDFs with record sensitivities. 
In this letter, we report new experimental studies on the SSVDFs
by using ISSCs and a SERF co-magnetometer \cite{2016yao, Yaodynamics2016,magnetic2016Yao,Smiciklas:2011}.

%=================================================================
%=================================================================
%=================================================================
%=================================================================
\paragraph{The SERF's Response to the SSVDFs--}
%=================================================================
%=================================================================
%=================================================================
SSVDFs to be studied  here are, 
following the notation in Ref. \cite{16Forms2006Dobrescu,leslie2014prospects},
$V_{6+7}$, $V_8$, $V_{15}$ and $V_{16}$.
For example, $V_{16}$ can be written as,
\begin{align}
\begin{split}
V_{16}=&- \frac{ f_{16}\hbar^2}{8\pi m_\mu c^2}
\big\{ 
(\hat{\boldsymbol\sigma_2}\cdot\mathbf{v})\left[ \hat{\boldsymbol\sigma_1}\cdot(\mathbf{v}\times\hat{\boldsymbol r}) \right]\\&+(\hat{\boldsymbol\sigma_1}\cdot\mathbf{v})\left[ \hat{\boldsymbol\sigma_2}\cdot(\mathbf{v}\times\hat{\boldsymbol r})\right]
\big\}
\left(\frac{1}{\lambda r}+\frac{1}{r^2}\right)e^{-r/\lambda},
\end{split}
\end{align}
where $f_{16}$ is a dimensionless coupling constant, 
 $\hat{\sigma}_1$, $\hat{\sigma}_2$  are the spins of the two particles respectively, and
$\mathbf{v}$ is the relative velocity between the two interacting fermions.
For this new interaction, the corresponding effective magnetic field $\mathbf{B}_{eff}$ experienced by the polarized spin due to the spin source can be deduced from 
$V_{16}=-\mathbf{\mu} \cdot \mathbf{B}_{eff}$, where $\mu$ is  the magnetic momentum of the probing particle. In a typical polarized noble gas experiment, 
the probe particles could be nuclei, e.g. $^{21}$Ne, or  valence electrons of alkali.

If this $\mathbf{B}_{eff}$ exists, 
the SERF's response can be estimated by the Bloch equations\cite{Kominis:2003},  % \cite{}????????????????????
\begin{equation}\label{eqn.pe}
\frac{\partial \mathbf{P}^e}{\partial t}=
\frac{\gamma_e}{Q(P^e)}\left[ \mathbf{B}_{eff}^e+\mathbf{B}+\lambda M^n \mathbf{P}^n+\mathbf{L} \right ]\times \mathbf{P}^e 
+ \frac{P_0^e\hat{\mathbf{z}}-\mathbf{P}^e}{T_e Q(P^e)},
\end{equation}
\begin{equation}\label{eqn.pn}
\frac{\partial \mathbf{P}^n}{\partial t}=
\gamma_n \left[ \mathbf{B}_{eff}^n+\mathbf{B}+\lambda M^e \mathbf{P}^e \right ]\times \mathbf{P}^n 
+ \frac{P_0^n\hat{\mathbf{z}}-\mathbf{P}^n}{\{T_{2n}, T_{2n}, T_{1n}\}},
\end{equation}
where $\mathbf{B}_{eff}^{e,n}$ are the effective magnetic fields due to the possible new SSVDFs coupling to the electron (or nucleon) spin;
$\mathbf{P}^{e,n}$  are the polarization of electron or nucleon respectively; 
 $\mathbf{B}$ is the external magnetic field;
 $T_e$, $T_{1n}$, and $T_{2n}$ are the electron spin's relaxation time,  
 nucleon spin's longitudinal and transverse relaxation times respectively;
$M^{e,n}$ are the magnetization associated with the electron or the nucleon spin;
$P_0^e$ ($P_0^n$) is the equilibrium polarization of the electron ( nucleon); 
$\mathbf{L}$ is the pumping light induced effective magnetic field experienced by the electron spin;
$Q(P^e)$ is the electron slow-down factor associated with the hyperfine interaction and spin-exchange collisions \cite{Kornack:2002}; 
and $\gamma_e$ ($\gamma_n$)  is the gyromagnetic ratio of the electron (nucleon).
It is worth noticing that the Eq.(\ref{eqn.pe}) and (\ref{eqn.pn}) are coupled together.
For example, if $B^{n}_{eff}= 0$, but   $B^{e}_{eff}\neq 0$,
the SERF still has nonzero output $S^{sim}(t)$.

\begin{figure}
\begin{center}
\includegraphics[width=9.cm]{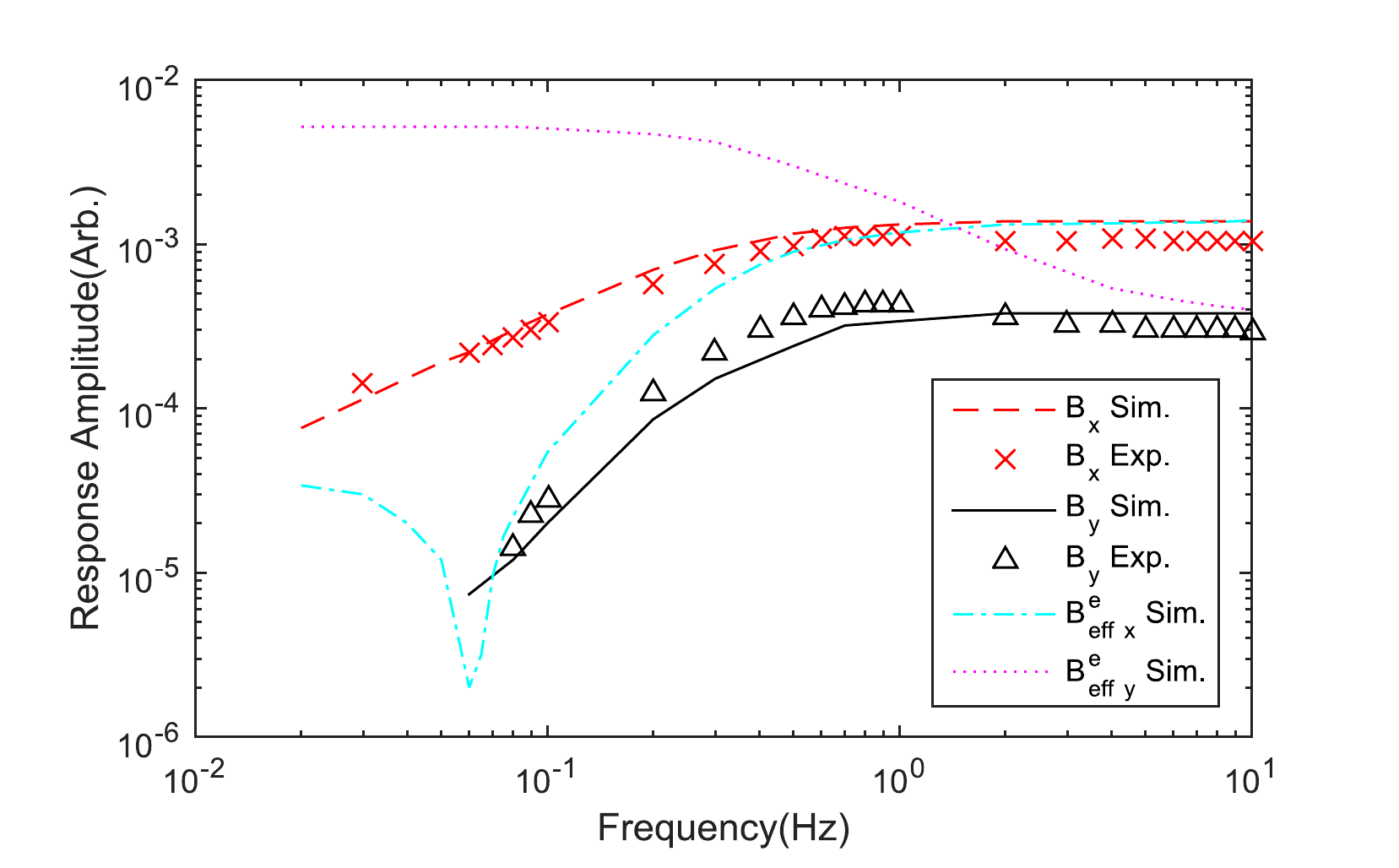}
\caption{
(Color online) 
SERF's response to magnetic fields. The red crosses and red dash line are experimental and simulation results for $B_x$ respectively, while the black triangles and black solid line are for $B_y$. The simulation results were obtained by solving Eq. \ref{eqn.pe}\&\ref{eqn.pn} numerically. The pink dot line and blue dash-dot line are simulations with assumptions that the exotic force only affected electrons. 
 }
\label{Fig.SERF.response} 
\end{center}
\end{figure}

By solving the equation set (\ref{eqn.pe}) and (\ref{eqn.pn}) numerically,
one can convert the SERF's response $\mathbf{P}^e(t)$ to a variable field 
$\mathbf{B}^{e,n}_{eff}$. The numerical results, together with the experimental measurements, are shown in Fig. \ref{Fig.SERF.response}.  
As shown in Fig. \ref{Fig.SERF.response}, the sensitivity of the co-magnetometer's response was frequency dependent. 

\begin{figure}
\begin{center}
\includegraphics[width=9.cm]{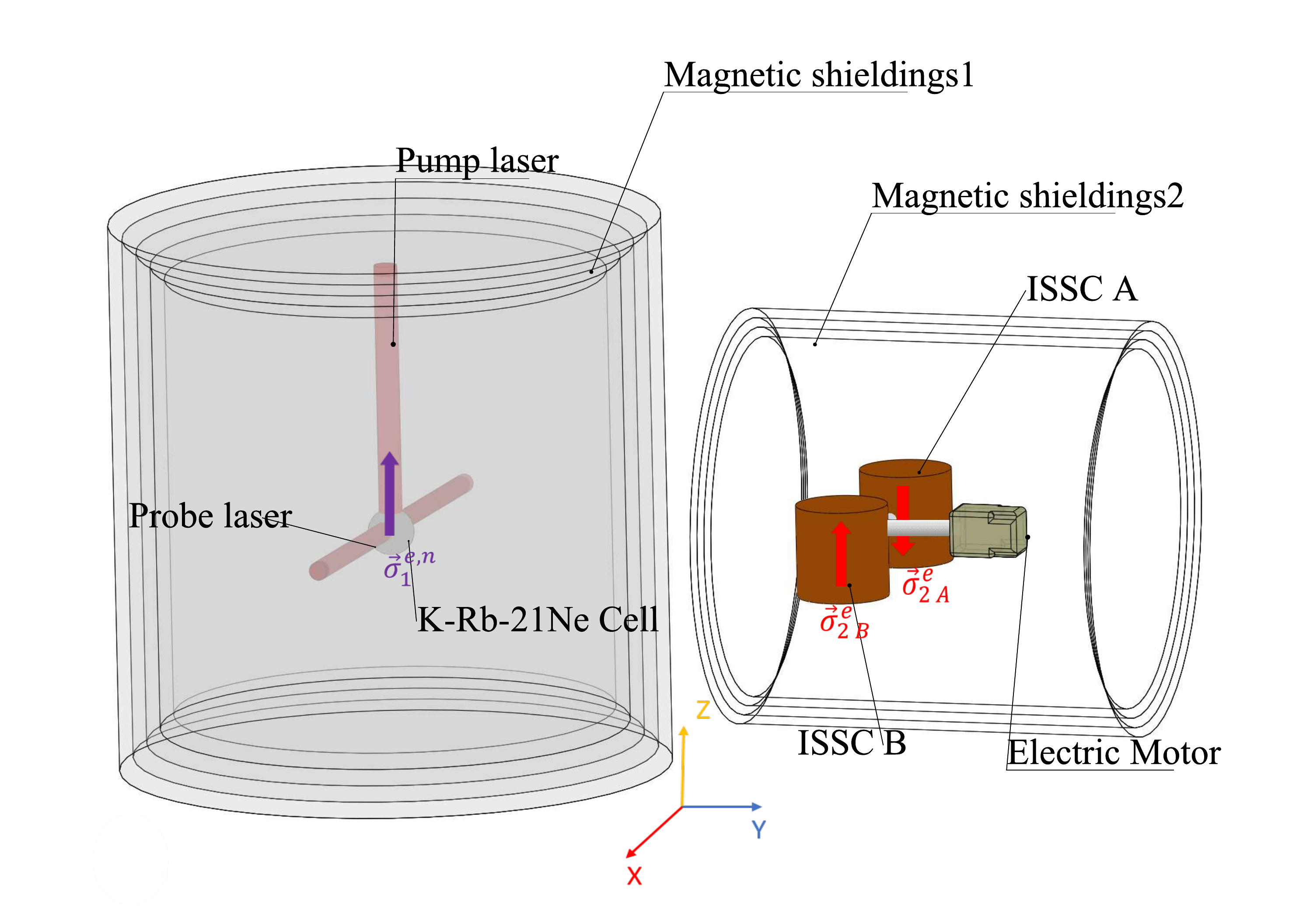}
\caption{
(Color online)The experimental setup. 
The left side was the SERF co-magnetometer.
%with polarized K and Rb electrons served as $\sigma_1$.
The two ISSC spin sources noted as 2A and 2B 
%served as $\sigma _2$. 
were driven by a servo motor, and they could rotate CW and CCW along y axis with a given frequency. 
}
\label{Fig.Exp.Setup} 
\end{center}
\end{figure}

%=================================================================
%=================================================================
%=================================================================
%=================================================================
%\section{Experimental Setup}
\paragraph{Experimental Setup--}
%=================================================================
%=================================================================
%=================================================================
The experiment was carried out at Beihang University, Beijing, China. 
The setup is shown in Fig. \ref{Fig.Exp.Setup} schematically. 
The left side is a SERF co-magnetometer. A detailed description of the device can be found in Ref. \cite{2016yao}.
A spherical aluminosilicate glass vapor cell with a diameter of 14 mm was located at the center of the SERF. 
It was filled with 3 bar of $^{21}$Ne gas (isotope enriched to 70$\%$ ), 53 mbar  of N${_2}$ gas, and a small amount of  K-Rb mixture. The mixture mole ratio was about 0.05 for the hybrid pumping purpose \cite{Smiciklas:2011}.
The cell was shielded by four layers of $\mu$-metal and a layer of 10-mm-thick ferrite \cite{Kornack:2007} magnetic field shielding to reduce the ambient magnetic field. 

As shown in Fig. \ref{Fig.Exp.Setup}, 
a linearly polarized probe laser  beam, which was modulated by a  50 kHz signal, passed through the cell, and 
its Faraday rotation angle was then measured by using photo-elastic modulation (PEM).
The signals from  photo-diodes were amplified by a lock-in amplifier, 
which had a reference frequency of 50 kHz, the same as the probe's modulation.
The lock-in output was then recorded by a data-acquisition system.

As shown at the right side in Fig. \ref{Fig.Exp.Setup} , there were two ISSCs, the electron spin sources. 
They were identical iron-shielded SmCo$_5$ (ISSC) magnets \cite{ji2017searching}.
Each ISSC had a cylindrical SmCo$_5$ magnet inside, which was covered by 3 layers of pure iron.
The magnets were cylindrical with diameter $40.00$ mm and height $40.00 $ mm. Thicknesses of the iron shielding layers were $15.00$ mm, $5.00 $ mm, and $5.00 $ mm respectively. 
The internal magnetic field of the SmCo$_5$ magnet was about 1 T.

%=================================================================
%=================================================================
%=================================================================
%=================================================================
%\subsection{Experimental Procedure}
%=================================================================
%=================================================================
%=================================================================

\begin{figure}
\begin{center}
\includegraphics[width=9cm]{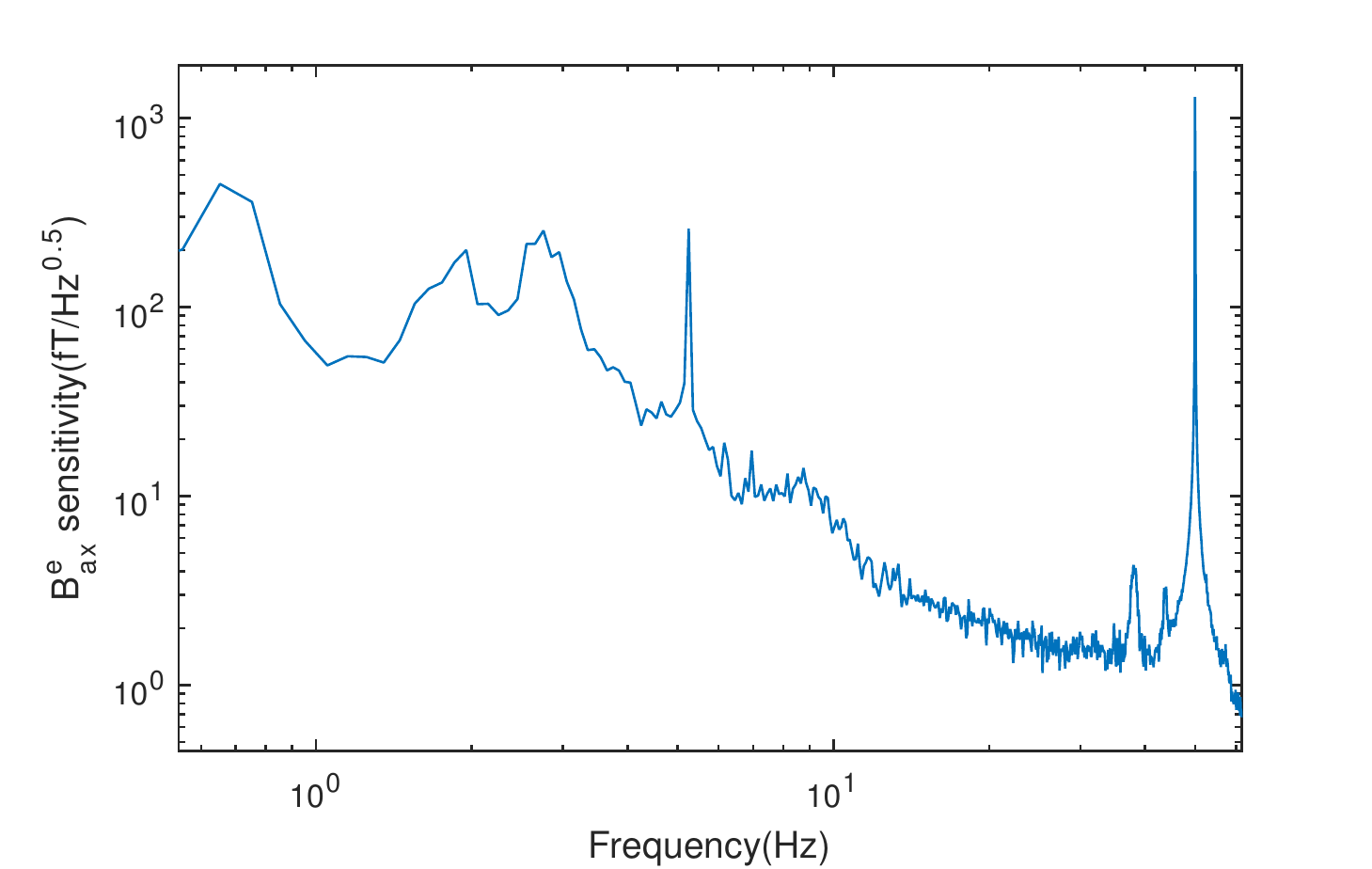}
\caption{
A typical power spectrum measured in the experiment. 
The ISSCs rotated in the frequency of 5.25 Hz, and the motor power system may cause the peak at 5.25 Hz in the spectrum.
The 50 Hz peak in the spectrum came from the power supply of the equipment, which was 50 Hz, 220V.
}

\label{fig.power.spectrum}  
\end{center}
\end{figure}

Driven by a servo motor, the ISSCs rotated with a frequency of $f_{0}=5.25$ Hz clockwise (CW) or counter-clockwise (CCW). 
Because the two ISSCs were mounted centrosymmetrically,
%and the driving motor was rotating with a frequency of 5.25 Hz,
the frequency of the possible SSVDF signals detected by the SERF was doubled, i.e. 10.5 Hz. This frequency was chosen due to the facts that the SERF co-magnetometer had relatively large responses to both $B^e_{eff, x}$ and $B^e_{eff, y}$ (Fig. \ref{Fig.SERF.response}), as well as relatively low noise level here (Fig. \ref{fig.power.spectrum}).
%, and the higher frequency, the higher velocity.
When rotating to a given angle, the ISSCs  could trigger an optoelectronic pulse,
and this signal was recorded by the data-acquisition system. 
This signal was used as the starting point of a new cycle for data analysis. 
Similar to the SERF, the ISSCs as well as the servo motor were both shielded by 4 layers of $\mu$-metal to further reduce possible magnetic field leakage from the ISSCs and servo motor.

%=================================================================
%=================================================================
%=================================================================
\paragraph{Data Analysis--}
%=================================================================
%=================================================================
%=================================================================
The experimental raw data was recorded as $\mathbf{S}^{exp}_{i,raw}(t_{j})$, where $i$ and $j$ mean the $j$-th point in the $i$-th cycle, $t_j=j*\Delta t$, and $ \Delta t$ is the data sampling period.
Then,  $\mathbf{S}^{exp}_{i,raw}(t_{j})$ was first transformed to frequency domain by using Fast Fourier Transformations (FFT). 
A typical SERF power spectrum is shown in Fig.\ref{fig.power.spectrum}.
Then Gaussian filters were applied to remove  the peaks corresponding to 5.25 and 50 Hz.
After that, the signals were transformed back to the time domain with inverse FFT. 
Furthermore,  DC components in $\mathbf{S}^{exp}_{i,raw}(t_{j})$ were also removed. After the steps above, the raw signals $\mathbf{S}^{exp}_{i,raw}(t_{j})$ were then transferred to $\mathbf{S}^{exp}_{i}(t_{j})$ for further analysis.

\begin{figure}
\begin{center}
\includegraphics[width=9.5cm,trim=0 0 0 0]{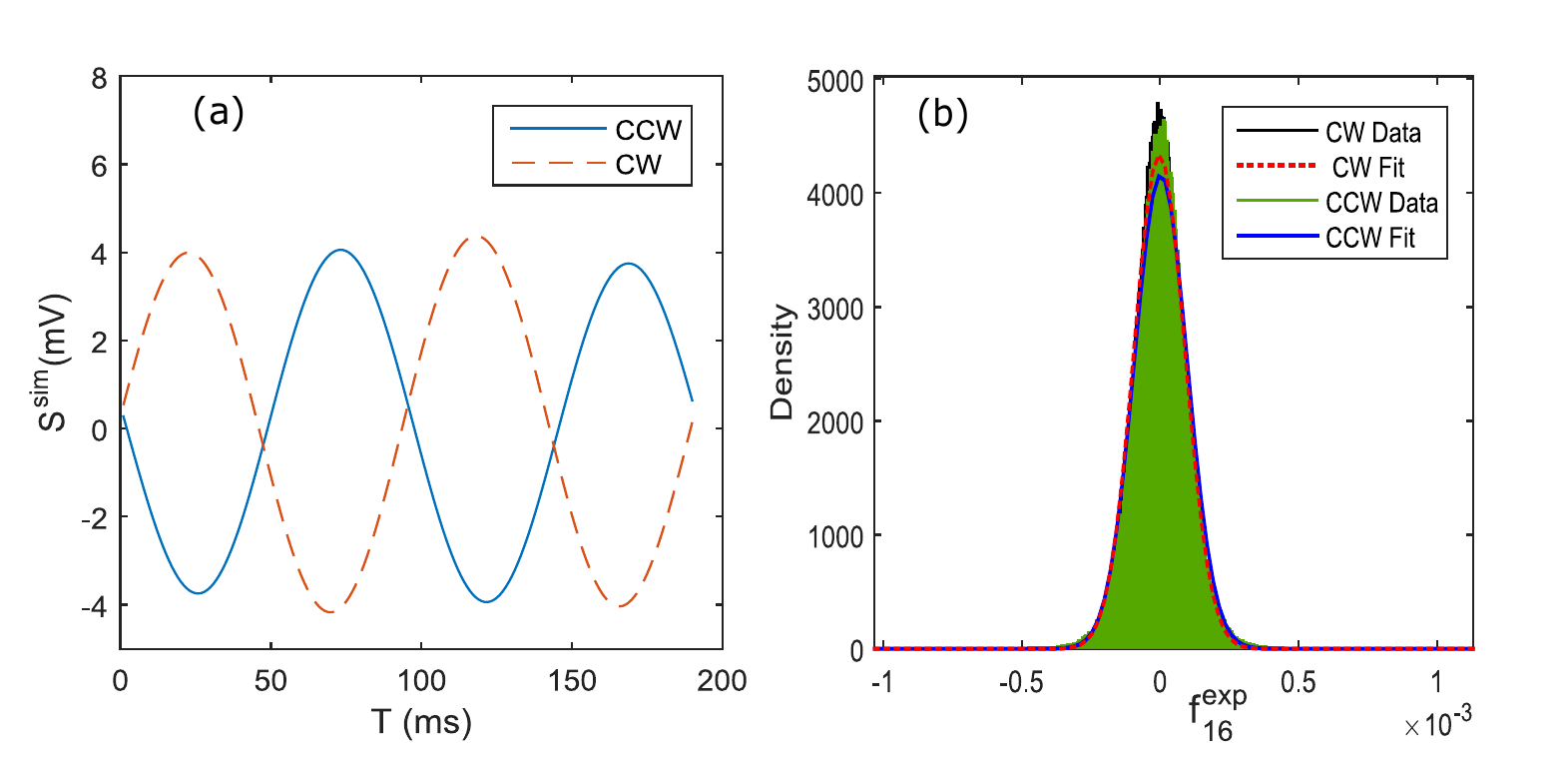}
\caption{(Color online)
(a) The simulated $\mathbf{S}^{sim}_{16}$ signals with $f^{(tn)}_{16}=1 \times 10^{-4}$, $\lambda=1000$ when motor rotates CW (blue solid line) and CCW (red dashed line). 
(b) The distribution of the $f_i^{exp}$ with $S^{ref}=S^{sim}_{16}$.
The black area represents the ISSCs rotating CW, and green area CCW.  
The red dashed line and blue solid line are their Gaussian fit respectively.
}

\label{fig.signal}
\end{center}
\end{figure}

Expected signals $S^{sim}(t)$ sensed by the SERF could be simulated by solving the equation sets (\ref{eqn.pe}) and (\ref{eqn.pn}) with the experimental parameters and a tentative coupling constant $f_{16}^{(tn)}$ in $V_{16}$ as inputs.
In the parameter space that we were interested in, i.e., $B_{eff}< 1 $ nT, 
$P_x^e(t)$ approximately linearly dependent on $B_{eff}$, and thereafter the coupling constant $f_{16}$.
The $S^{sim}(t)$ then linearly depended on $B_{eff}$, i.e. $S^{sim}(t) \simeq  \kappa f_{16}^{(tn)}\,B_{eff}$,
where  $\kappa$ is the calibration constant, which  was measured to be $110\pm 5$ V/nT.
The input $B_{eff}$ for solving Eq. \ref{eqn.pe}\&\ref{eqn.pn}
were simulated by the finite element analysis method\cite{ji2017searching}.
Two examples of  simulated signals for $V_{16}$, the $\mathbf{S}^{sim}_{16}$,  with motor rotating CW and CCW are shown in Fig. \ref{fig.signal}(a). 

The experimental signals $\mathbf{S}^{exp}_{i}(t_{j})$ were then compared with the simulated ones $S^{sim}(t)$.
A cosine similarity score $k_i$ was used to weigh the similarity between  $\mathbf{S}^{exp}_i$ and a given reference signal $\mathbf{S}^{ref}(t)$, 
which can be written as \cite{PhysRevE.92.042927},
\begin{equation}\label{eq.ki}
k_i
\equiv\frac{\sum_j \mathbf{S}^{ref}(t_j)\cdot {\mathbf{S}}_i^{exp}(t_j)}
{\sqrt{\sum_j \left[\mathbf{S}^{ref}(t_j)\right]^2}
\sqrt{\sum_j \left[{\mathbf{S}_i}^{exp}(t_j)\right]^2}
}.
\end{equation}

The coupling constant in $V_{16}$ measured experimentally in $i$-th cycle, $f^{exp}_{i,16}$, can be written as,
\begin{equation}
f^{exp}_{i,16}=k_i f_{16}^{(tn)}\, \sqrt{\frac{\sum_j \left[\mathbf{S}^{exp}_i(t_j)\right]^2}
{\sum_j \left[\mathbf{S}^{ref}(t_j)\right]^2}}.
\end{equation}
Distributions of $f^{exp}_{i,16}$ are shown in Fig. \ref{fig.signal}(b). They agree with Gaussian shapes well.

The final experimentally measured coupling constant $f^{exp}_{16}$ was obtained by averaging all rotating cycles including CW and CCW, i.e.
\begin{equation}\label{eq.average}
 f^{exp}_{16} =
\frac{\langle f^{exp}_{i,16}\rangle_+ 
+
\langle f^{exp}_{i,16}\rangle_-}
{2}
,
\end{equation}
where $\langle f^{exp}_{i,16}\rangle_+=\frac{1}{n}\sum_{i=1}^{n} f^{exp}_{i,16}$
is the average over the CW cycles, 
and $\langle f^{exp}_{i,16}\rangle_-$, the CCW cycles.

\begin{figure*}[ht]
\begin{center}
 \includegraphics[width=18.cm,height=10 cm,trim=0 0 0 0]{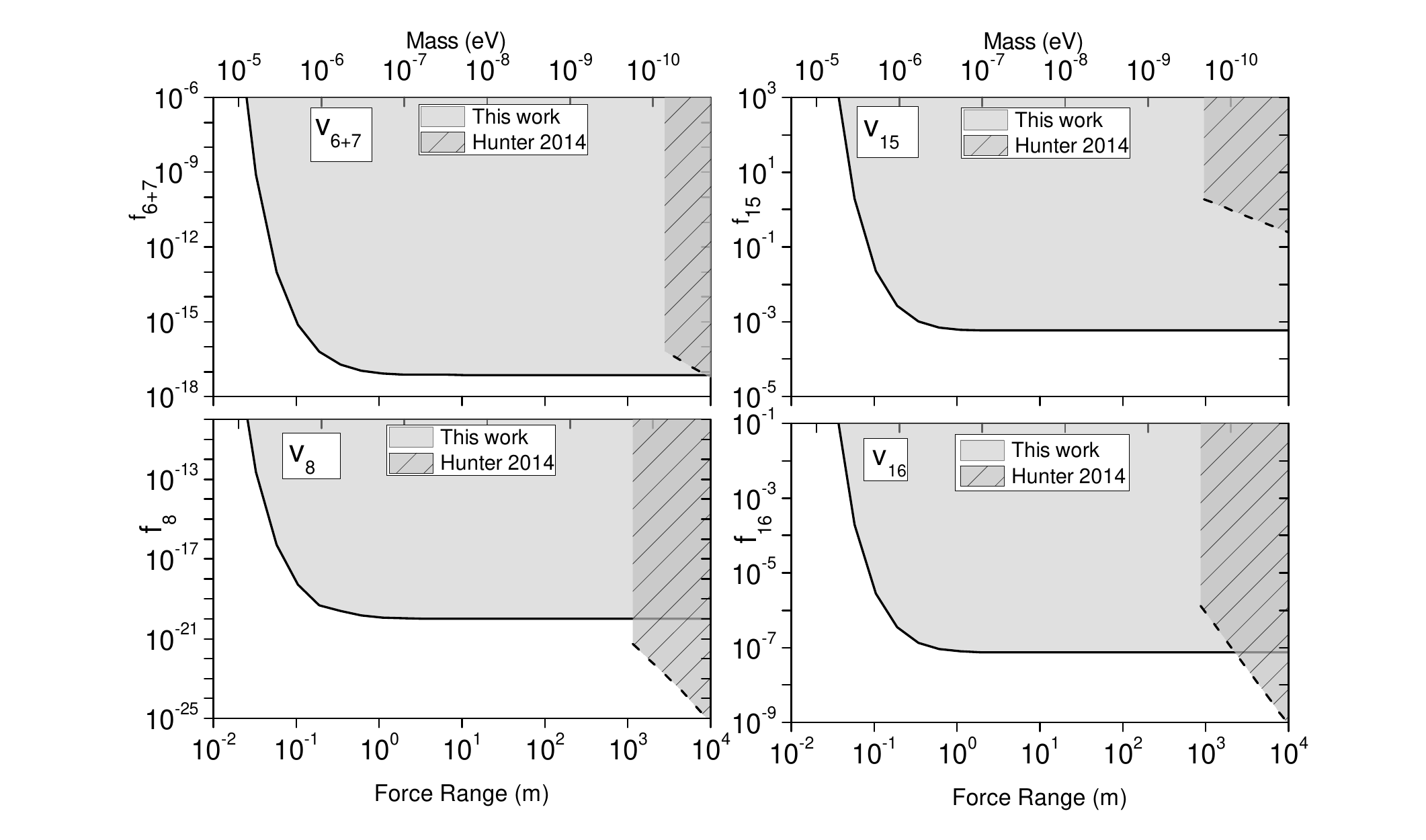}
\caption{
Limits on the SSVDFs' coupling constants between two electrons measured in this work, and comparison with those in literature. The "Hunter2014" comes from Ref. \cite{hunter2014using}, in which polarized geo-electrons were used. 
}
 \label{fig.results}  
\end{center}
\end{figure*}

%=================================================================
%=================================================================
%=================================================================
%\subsection{Data Analysis}
\paragraph{Results and Discussion--}
%=================================================================
%=================================================================
%=================================================================

The other terms of SSVDFs \cite{16Forms2006Dobrescu,leslie2014prospects} were analyzed by the same method. These interactions were:
\begin{align}
V_{\rm 6+7}=&\frac{ -f_{6+7} \hbar^2}{4\pi m_\mu c}
(\hat{\boldsymbol\sigma_1} \cdot \mathbf{v})(\hat{\boldsymbol\sigma}_2\cdot\hat{\boldsymbol r})
\left(\ \frac{1}{\lambda r}+\frac{1}{r^2} \right) e^{-r/\lambda},\\ 
%V_{14}=&\frac{f_{14}\hbar }{4\pi}\left[(\hat{\boldsymbol\sigma_1}\times\hat{\boldsymbol\sigma_2})\cdot\mathbf{v}\right]\left(\frac{1}{r}\right)e^{-r/\lambda},\\
\begin{split}
 V_{8}&=\frac{ f_8 \hbar}{4 \pi c}
 (\hat{\boldsymbol\sigma_1} \cdot\mathbf{v})
 (\hat{\boldsymbol\sigma_2} \cdot\mathbf{v})
\frac{e^{-r/\lambda}}{r}
 \end{split},\\
\begin{split}
V_{15}=& \frac{-f_{15}\hbar^3 }{8\pi m_1m_2 c^2}
\big\{
(\hat{\boldsymbol\sigma_2}\cdot\hat{\boldsymbol r}) \left[ \hat{\boldsymbol\sigma_1}\cdot(\mathbf{v}\times\hat{\boldsymbol r}) \right]+(\hat{\boldsymbol\sigma_1}\cdot\hat{\boldsymbol r})\\&\left[ \hat{\boldsymbol\sigma_2}\cdot(\mathbf{v}\times\hat{\boldsymbol r})\right]
\big\}
\left(\frac{1}{\lambda^2r }+\frac{3}{\lambda r^2} + \frac{3}{r^3}\right)e^{-r/\lambda}.
 \end{split}
\end{align}

%=================================================================
%=================================================================
%=================================================================
%\subsection{Data Analysis}
%\paragraph{Discussion--}
%=================================================================
%=================================================================
%=================================================================

The parameters of the setup and their errors are shown in Tab. \ref{parameters.tab}. 
Considering these errors, together with the statistical error, 
the  constraints  on the SSVDFs between two electrons could be set. The results are  shown in Fig. \ref{fig.results}. 
The gray areas  are excluded with 95\% confidence level. 
For $V_{6+7}$,  $V_{8}$, $V_{15}$, and $V_{16}$, our experiment can set up new record limits at the range of 5 cm -- 1 km.  
Especially for $V_{15}$, our result is over 3 orders of magnitude better than \citep{hunter2014using} in force range between 5 cm and 1 km. 

The error budget for $f^{exp}_{i,16}$ 
at $\lambda=1.1$ m
is shown in Tab. \ref{parameters.tab}.
The major systematic error came from
the cross-talking between the servo motor power system and the SERF system.  
The 5.25 Hz peak shown in Fig. \ref{fig.power.spectrum} might come from cross-talking effect.  
However, the major frequency considered here was 10.5 Hz, whose amplitude was about 40 times smaller than 5.25 Hz.
The secondary harmonics of 5.25 Hz could also contribute to systematic error. 
In fact, the correlation between 5.25 Hz and 10.5 Hz could be calculated 
by applying $\mathbf{S}^{ref}_A(t_j)=\sin[5.25t_j]$ or $\mathbf{S}^{ref}_B(t_j)=\sin[10.5t_j]$ to Eq.(\ref{eq.ki}).
A correlation between 5.25 and 10.5 Hz was indeed found this way, which confirmed the cross-talking effect.
The cross-talking was the dominant effect in our experiment.

Another major consideration was the magnetic leakage from the ISSCs.
With the iron shielding, at a distance of 20 cm away from the ISSC's mass center, 
its residual magnetic field was measured to be $< 10$ mG. 
The  magnetic shielding factors for the mu-metals outside the ISSCs were measured to be $>10^{6}$, and shielding for SERF magnetometer,  $>2\times 10^{6}$. 
Considering all factors together, we conservatively expect the magnetic leakage from the ISSCs to SERF's center to be smaller than $10^{-2}$ aT, which was insignificant in regards to the error budget.

It is worth pointing out that only the errors of the parameters when doing the calculation of $f^{sim}$ as well as the statistic uncertainty could affect the limit curves drawn in Fig. \ref{fig.results}.
The magnetic field leakage and cross-talking were not subtracted in these plots.

\begin{table}[!h]
\begin{ruledtabular}
\caption {Input parameters for the FEA simulation and their error contribution to the final uncertainties. The origin of coordinates was at the center of the pumping cell.} 
\label{parameters.tab}
\begin{tabular}{c c c c}   
Parameter & Value & $\Delta f^{exp}_{16} (\times 10^{-8})$
\footnote{The contribution  to the error budget of $V_{16}$ 
at $\lambda=1.1$ m 
}  \\
\hline
ISSC net spin ($\times10^{24}$) &$1.75\pm 0.21 $ & $^{+ 0.34}_{-0.32}$\\
Position of ISSCs y(m)& $-0.624\pm 0.005$ & $^{+ 0.13}_{-0.12} $\\ 
Position of ISSCs z(m)& $0.278\pm 0.005$  & $\pm 0.04$\\
%SmCo$_5$ magnetization(T)  &$1 \pm 0.05$  & \ \\
D between 2 ISSCs(m) & $0.251\pm 0.001$   & $\pm 0.03$\\
Rotating frequency(Hz) &$5.250\pm 0.001$  & $ < 0.001$\\
Calib. const. $\kappa$ (V/nT) & $330\pm20$ & $\pm 0.23$ \\
phase uncertainty ($\deg$) &$\pm 5 $& $\pm 0.27$\\
\hline
Final $f^{exp}_{16} ( \times 10^{-8})$ & $4.0$
 & $\pm 1.9\ (statistic)$\\

$(\lambda=1.1 m)$ &  & $\pm 0.5
\footnote{
Error contribution from the uncertainties of the parameters listed above.
} 
$ 
\end{tabular} 

\end{ruledtabular}
\end{table}

%====Summary===============
%====Summary===============
%\section{Summary}
\paragraph{Summary--}
%====Summary===============
%====Summary===============
%\paragraph{Summary--}

In summary,  
by using  specially designed iron-shielded SmCo$_5$ permanent magnets, a high  electron spin density source of about $ 1.7\times 10^{21}$ cm$^{-3}$  has been achieved, while still keeping its magnetic leakage down to about mG level.
The similarity analysis have been proved to be successful, 
which gives a boost to the detecting sensitivities. 
With help from the high spin density, the high sensitive SERF co-magnetometer, and the similarity analysis, 
new constraints on possible new exotic potentials of $V_{6+7}$, $V_8$, $V_{15}$, and $V_{16}$ 
were derived for  force range of 5 cm -- 1 km. To the best of our knowledge, it is the first time these results have beem attained.
By dedication to improving the SERF sensitivities, and reducing the crossing-talking effect,
a higher sensitivity by a factor of over 1000 is expected in future studies with a similar experimental setup.

\begin{acknowledgments}
This work is supported by Tsinghua University  Initiative Scientific Research Program,
and the National Natural Science Foundation of China (NSFC) under Grant No. 11375114, 91636103, and 11675152. This work is also supported by the Key Programs of the NSFC under Grant No. 61227902.
%H.Y. acknowledges support from the NSFC under Grant .  
\end{acknowledgments}

%\end{CJK*}
%\bibliographystyle{apsrev4}
%\bibliography{5thForce}
%

\end{document}